\documentclass[12pt]{iopart}
\usepackage[final]{graphicx}
\begin{document}

\title[Topologically close-packed phases in binary transition-metal compounds]
      {Topologically close-packed phases in binary transition-metal compounds: 
       matching high-throughput ab-initio calculations to an empirical structure-map}

\author{T~Hammerschmidt$^{1,2}$, AF~Bialon$^{1}$, DG~Pettifor$^{2}$ and R~Drautz$^{1,2}$}

\ead{thomas.hammerschmidt@rub.de}

\address{$^1$ ICAMS, Ruhr-Universit\"at Bochum, Bochum, Germany, \\
         $^2$ Department of Materials, University of Oxford, Oxford, United Kingdom}

%\date{\today}

\begin{abstract}

In steels and single-crystal superalloys the control of the formation of topologically close-packed (TCP) phases
is critical for the performance of the material. The structural stability of TCP phases in multi-component 
transition-metal alloys may be rationalised in terms of the average valence-electron count $\bar{N}$ and the composition-dependent relative 
volume-difference $\overline{\Delta V/V}$.
We elucidate the interplay of these factors by comparing density-functional theory calculations to an empirical structure map
based on experimental data. In particular, we calculate the heat of formation for the TCP phases A15, C14, 
C15, C36, $\chi$, $\mu$, and $\sigma$ for all possible binary occupations of the Wyckoff positions. 
We discuss the isovalent systems V/Nb-Ta to highlight the role of atomic-size difference and observe the expected stabilisation of 
C14/C15/C36/$\mu$ by $\overline{\Delta V/V}$ at $\Delta N=0$ in V-Ta. 
In the systems V/Nb-Re, we focus on the well-known trend of A15$\rightarrow \sigma \rightarrow \chi$ 
stability with increasing $\bar{N}$ and show that the influence of $\overline{\Delta V/V}$ is too weak to 
stabilise C14/C15/C36/$\mu$ in Nb-Re. 
As an example for a significant influence of both $\bar{N}$ and $\overline{\Delta V/V}$, we also consider the  
systems Cr/Mo-Co. Here the sequence A15$\rightarrow \sigma \rightarrow \chi$ is observed in both systems but 
in Mo-Co the large size-mismatch stabilises C14/C15/C36/$\mu$. 
We also include V/Nb-Co that cover the entire valence range of TCP stability and also show the stabilisation 
of C14/C15/C36/$\mu$. Moreover, the combination of a large volume difference with a large mismatch in valence-electron
count reduces the stability of the A15/$\sigma$/$\chi$ phases in Nb-Co as compared to V-Co.
By comparison to non-magnetic calculations we also find that magnetism is of minor importance for the
structural stability of TCP phases in Cr/Mo-Co and in V/Nb-Co.
\end{abstract}

%\pacs{71.20.Lp,31.15.A-,71.15.Nc}
% 31.15.A-  Ab initio calculations 
% 71.20.Lp  Intermetallic compounds 
% 71.15.Nc  Total energy and cohesive energy calculations 

\maketitle

\section{Introduction}

The topologically close-packed (TCP) phases are a class of crystal structures that are observed in 
many intermetallic compounds~\cite{Sinha-73,Hartsough-74,Stein-04,Joubert-08,Joubert-09,Seiser-11-1}.
Depending on the application, the goal of materials design with regard to TCP phases is either to avoid 
or to enforce and control their precipitation.
In Ni-base single-crystal superalloys for high-temperature applications, the formation of TCP phases
leads to a degradation of mechanical properties~\cite{Rae-01}. The same detrimental effect can be 
expected for the more recently discussed Co-base single-crystal superalloys~\cite{Sato-06}. 
In plasma-facing alloys based on refractory metals for fusion reactors, the formation of TCP phases following nuclear 
transmutations~\cite{Nemoto-04} may cause mechanical failure due to internal stresses on the brittle TCP 
phase precipitates~\cite{Cottrell-04}. 
In precipitate-hardened steels, the TCP phases are used as obstacles to dislocation movement 
for improved creep strength~\cite{Aghajani-09}. 

Many of the TCP phases have been investigated earlier with respect to general trends in structural 
stability, particularly the A15 phase~\cite{Hartsough-74,Turchi-83,Turchi-92,Moriarty-94}, the Laves 
phases~\cite{Ohta-90,Zhu-02,Stein-04}, the $\chi$ phase~\cite{Joubert-09} and the $\sigma$ phase~\cite{Joubert-08}. 
The prediction of structural stability of TCP phases was initially effectively based on the average 
number of valence electrons as the only parameter~\cite{Rideout-51,Boesch-64,Morinaga-84,Watson-84}. 
In a recent structure-map~\cite{Seiser-11-1} a second parameter was added that accounts for 
the known influence of atomic-size differences~\cite{Laves-42}:
The resulting structure map for TCP phases~\cite{Seiser-11-1} shows regions of TCP phase occurrence in terms 
of the average valence-electron count $\bar{N}$ and the composition-dependent relative 
volume-difference $\overline{\Delta V/V}$.
The trends in structural stability as displayed by the structure map can be rationalized and understood in a 
tight-binding analysis of the electronic structure and a local description of bond formation using analytic 
bond-order potentials~\cite{Seiser-11-2}. The valence electron concentration was shown to stabilize the A15,
$\sigma$ , and $\chi$ phases at approximately half full band but to destabilize the $\mu$ and Laves phases. 
A significant relative size difference between the constituent atoms of an alloy is required to 
stabilize the $\mu$ and Laves phases. The fourth-moment contribution to the bond energy suffices to explain 
the separation of A15, $\sigma$, and $\chi$ phases from the $\mu$ and Laves phases. The differences in the 
fourth-moment contribution for the structures are related to differences in the bimodality of the electronic
density of states which are caused mainly by distortions of the local coordination polyhedra as compared to 
the ideal Frank-Kasper polyhedra.
The existence of such regions of structural stability has been shown earlier for p-d bonded AB compounds with a 
tight-binding model~\cite{Pettifor-84,Pettifor-86-2}. There, the relative stability of NaCl, CsCl, NiAs, MnP, CrB
and FeB could be explained in terms of an atomic size factor, the bandfilling and the p-d energy level separation. 
While the influence of $\bar{N}$ on the relative structural stability of TCP phases was also observed in 
atomistic calculations with density-functional theory~\cite{Berne-99,Berne-02} and approximate electronic 
structure methods~\cite{Seiser-11-1,Turchi-91}, it required the introduction of $\overline{\Delta V/V}$ 
to achieve a separation of $\mu$/C14/C15/C36 from A15/$\sigma$/$\chi$ in the structure map.
Density-functional theory (DFT) calculations were also used to determine the temperature-dependent occupancies of Wyckoff positions for 
individual phases, e.g., for the $\sigma$ phase in Re-W~\cite{Berne-01, Fries-02} and Cr-Ru/Os~\cite{Sluiter-09}, 
and for the $\mu$-phase in Ni-Nb~\cite{Sluiter-03,Dupin-06}. 
More recent DFT studies included further competing crystal structures and demonstrated the importance of entropic 
contributions to the free energy, e.g., of configurational entropy for Mo/W-Re~\cite{Crivello-10-1}, of vibrational 
entropy for disordered Re-W~\cite{Chouhan-12}, of vibrational entropy~\cite{Dubiel-10} and magnetic entropy for Fe-Cr~\cite{Pavlu-10}. 

Here, we systematically compare the predictions of the structure map for TCP phases~\cite{Seiser-11-1} 
to high-throughput DFT calculations for selected binary transition-metal (TM) compounds. This extends our previous work on 
understanding TCP stability at different levels of the electronic structure~\cite{Seiser-11-2} to a broader set of binary 
TM compounds including magnetic systems.
In Sec.~\ref{sec:TCP} we give a brief review of the structural stability of TCP phases and introduce the 
investigated binary systems in the light of the TCP structure map. In Sec.~\ref{sec:methodology} we present 
some technical details of our high-throughput environment for the DFT calculations. In Sec.~\ref{sec:results} we 
discuss our DFT results in detail and compare them to the predictions of the empirical structure map. We
conclude our findings in Sec~\ref{sec:conclusion}.

\section{Relevant compound phases}
\label{sec:TCP}

\subsection{Investigated crystal structures}

The topologically close-packed phases are a class of crystal structures that consist of regular Frank-Kasper polyhedra~\cite{Frank-58,Frank-59}. 
A detailed review and classification of these phases was given by 
Sinha~\cite{Sinha-73}. In our calculations we focus on TCP phases
that we identified previously~\cite{Hammerschmidt-08-2,Seiser-11-1, Seiser-11-2} as 
representatives of two groups of TCP phases: the A15, $\sigma$, and $\chi$ phase with their
stability being dominated by average valence-electron count, and the $\mu$ phase and the Laves phases 
(C14, C15, and C36) which are mostly stabilised by atomic-size differences. For each of these
TCP phases, we considered all possible occupations of the inequivalent lattice sites in the unit cell. 
In a binary system this results in $2^5=32$ possible stoichiometries for $\sigma$, $\mu$ and C36, 
in 16 for $\chi$, in 8 for C14 and in 4 for A15 and C15 phases. We excluded the TCP phases M, R, P 
and $\delta$ due to the vast number of possible stoichiometries in binary systems. We also excluded the 
recently proposed Nb$_2$Co$_7$ phase~\cite{Leineweber-12} that was observed only in a very small range of
narrow composition in the Nb-Co system. The basic crystallographic information and the 
number of inequivalent sites in the investigated phases is given in Tab.~\ref{tab:prototypes}. 
\begin{table}
\caption{\label{tab:prototypes} Strukturbericht designation, prototype and Pearson symbol of the TCP phases 
         A15, $\sigma$, $\chi$, $\mu$, C14, C15, and C36. The fifth column indicates the 
         number of inequivalent lattice sites in the unit cell.}
\begin{indented}
\item[]\begin{tabular}{@{}llll}
\br
Strukturbericht         & prototype          & Pearson       & inequivalent sites  \\
\mr
A15                     & Cr$_3$Si/$\beta$-W & {\it cP}8     & 2         \\
D8$_b$/A$_b$ ($\sigma$) & CrFe/$\beta$-U     & {\it tP}30    & 5         \\
A12 ($\chi$)            & $\alpha$-Mn        & {\it cI}58    & 4         \\
D8$_5$ ($\mu$)          & W$_6$Fe$_7$        & {\it hR}13    & 5         \\
C14 (Laves)             & MgZn$_2$           & {\it hP}12    & 3         \\
C15 (Laves)             & MgCu$_2$           & {\it cF}24    & 2         \\
C36 (Laves)             & MgNi$_2$           & {\it hP}24    & 5         \\
\br
\end{tabular}
\end{indented}
\end{table}
Similar to our stoichiometry sampling of the TCP phases, we also included ordered solid-solutions of bcc (e.g. B2, B11, B32, C11$_b$, DO$_3$, D0$_{23}$), 
fcc (e.g. B19, D0$_{22}$, D1$_a$, L1$_0$, L1$_1$, L1$_2$, L6$_0$), and 
hcp (e.g. B8$_1$, B8$_2$, B19, B35, D0$_{19}$) structures. This set of structures has been observed experimentally in the investigated 
binary systems or was suggested earlier for bcc/fcc~\cite{Cacciamani-97} and hcp~\cite{Hart-09} based structures. 
We also included structures that have been predicted to be stable in high-throughput DFT calculations of 
ordered binary alloys based on Re~\cite{Levy-10} or Ru~\cite{Jahnatek-11}. We also cover all phases that appear in the 
experimental phase diagrams of the investigated systems, except for Nb$_2$Co$_7$. 

\subsection{Investigated TM binary systems}

\begin{figure}[htb]
\begin{center}
\includegraphics[width=0.7\textwidth]{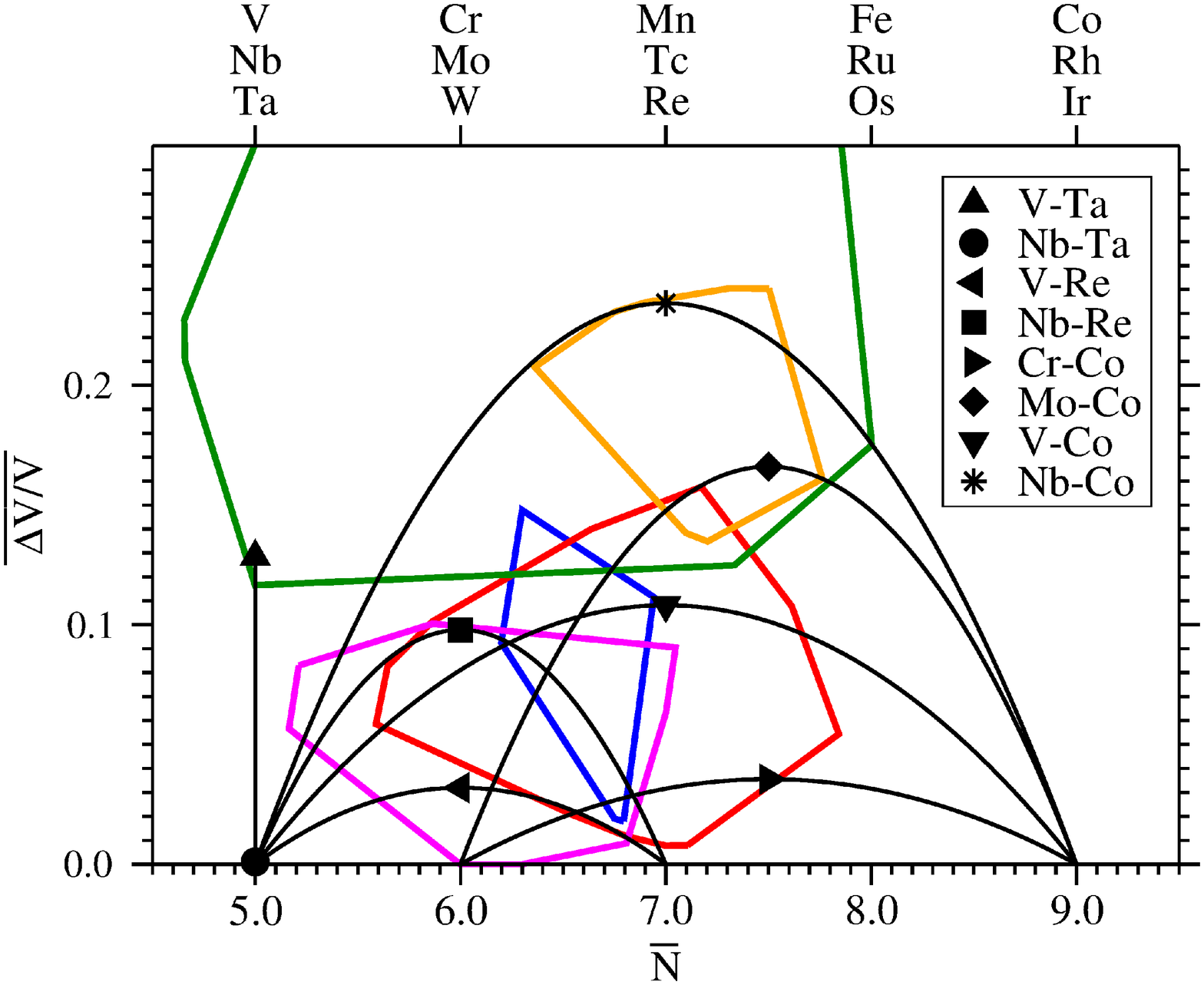}\\
\includegraphics[width=0.7\textwidth]{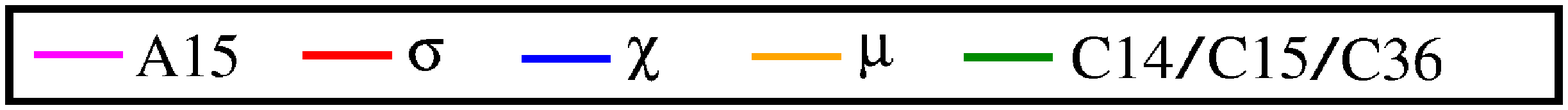}
\caption{The expected TCP phase stability is determined by intersecting the polygonal TCP regions of the 
structure map~\cite{Seiser-11-1} with the parabola that corresponds to varying chemical composition in 
a given binary TM compound. The different parabola correspond to the binary systems investigated in 
this work. The coloured polygonal areas indicate the TCP phases A15 (purple), $\sigma$ (red) $\chi$ (blue),
$\mu$ (orange), and C14/C15/C36 (green).} 
\label{fig:strucmap}
\end{center}
\end{figure}

We determined the stability of the aforementioned TCP phases in several binary TM compounds.
Our choice of TM binary systems is based on a recently devised structure map~\cite{Seiser-11-1} 
that rationalises the experimentally observed TCP phases by the average valence-electron 
count $\bar{N}$ and the composition-dependent relative volume-difference $\overline{\Delta V/V}$. 
We chose the systems V/Nb-Ta, V/Nb-Re and Cr/Mo/V/Nb-Co that cover the
central features of the empirical structure map. For each system we compiled the valence-electron
difference $\Delta N$, the relative volume-difference $\overline{\Delta V/V}$ at 1:1 composition and the
phases reported in experimental phase-diagrams~\cite{Landolt} in Tab.~\ref{tab:systems}. 
$\overline{\Delta V/V}$ is based on the metallic radii given in Ref.~\cite{Greenwood-book}. 
\begin{table}
\caption{\label{tab:systems} Difference in average valence-electron count $\Delta{N}$, relative 
volume-difference $\overline{\Delta V/V}$ based on metallic radii~\cite{Greenwood-book} of the 
investigated binary systems and experimentally observed phases~\cite{Landolt} including recent
reassessments for V-Ta~\cite{Pavlu-11} and Nb-Co~\cite{Stein-08}. Experimentally observed 
metastable phases are given in brackets. The valence-electron count of the elements, 
$N_A$ and $N_B$, are shown in Fig.~\ref{fig:strucmap}.}
\begin{indented}
\item[]\begin{tabular}{@{}cccc}
\br
A-B    & $\Delta N$  & $\overline{\Delta V/V}$ & sequence of phases A$\rightarrow$B  \\
\mr
V-Ta      & 0    & 0.128     & A2, C14, C15, A2          \\
Nb-Ta     & 0    & 0.001     & A2, A2                    \\
V-Re      & 2    & 0.032     & A2, A15, $\sigma$, A3     \\
Nb-Re     & 2    & 0.098     & A2, $\sigma$, $\chi$, A3  \\
Cr-Co     & 3    & 0.036     & A2, $\sigma$, (D0$_{19}$), (Co$_4$Cr-A3), A3   \\
Mo-Co     & 3    & 0.166     & A2, $\sigma$, $\mu$, D0$_{19}$, (Co$_4$Mo-A3), A3 \\
V-Co      & 4    & 0.108     & A2, A15, $\sigma$, D0$_{19}$$\rightarrow$L1$_2$$^{\mathrm{(a)}}$, A3  \\
Nb-Co     & 4    & 0.234     & A2, $\mu$, C14, C15, C36, Nb$_2$Co$_7$, L1$_2$$^{\mathrm{(b)}}$, A3 \\
\br
\end{tabular}
\item[(a) low temperature $\rightarrow$ high temperature transition]
\item[(b) precipitate only]
\end{indented}
\end{table}
We reproduce the TCP structure map of 
Ref.~\cite{Seiser-11-1} in Fig.~\ref{fig:strucmap} and highlight our set of TM binary 
systems. The variation of chemical composition for each TM binary system with elements $i$ and $j$ leads to 
a parabola in the structure map according to
\begin{equation}
\overline{\Delta V/V} = \sum\limits_{i,j} c_i c_j | V_i - V_j | /\left[ \left( V_i + V_j \right) /2 \right] \, ,
\end{equation}
as a function of valence-electron count $\bar{N}$.
The values of $\overline{\Delta V/V}$ at 1:1 composition and the range $N_A$ to $N_B$ of $\overline{N}$ 
are given in Tab.~\ref{tab:systems}.
For example for Nb-Re with respective atomic radii~\cite{Greenwood-book}  of 0.1468~\AA\, and 0.1375~\AA\,, 
the volume difference normalised to the average volume, $|Vi-Vj| / /\left[ \left( Vi+Vj \right) /2 \right]$, 
takes a value of 0.1957. At 1:1 ratio, i.e. at $c_i$=$c_j$=0.5, the value of $\overline{\Delta V/V}$ is 0.09786 
by summation over $i$-$j$=Nb-Re and Re-Nb.
In comparison to the experimental phase-diagrams~\cite{Landolt,Stein-08}, we find that for all binary 
systems, the experimentally observed TCP phases are also apparent in the structure map. 
In particular, the structure map correctly indicates the absence of TCP phases in Nb-Ta, the formation
of Laves or $\mu$ in V-Ta, Mo-Co and Nb-Co, as well as the formation of A15, $\sigma$, or $\chi$ in V-Re,
Nb-Re, Cr-Co, V-Co and Mo-Co.
Given the similarity of the phases within the two groups~\cite{Seiser-11-2} of A15/$\sigma$/$\chi$ and Laves/$\mu$, 
the structure map is apparently a useful tool to predict the maximum set of TCP phases that are to be expected 
and hence need to be considered in any attempt to determine the relative structural stability, e.g., by DFT 
calculations as shown in this work. 

\section{Ab-initio methodology}
\label{sec:methodology}

\subsection{Heat of formation}

The results presented in the following were determined by self-consistent total-energy calculation 
using density-functional theory. We used the VASP code~\cite{Kresse-94,Kresse-96-1,Kresse-96-2} 
with projector augmented wave pseudo-potentials~\cite{Bloechl-94} and the local-density approximation~\cite{Perdew-81}
to the exchange-correlation functional. We carried out convergence tests of the 
plane-wave cutoff and the Monkhorst-Pack $\mathbf{k}$-point mesh~\cite{Monkhorst-76} to converge 
the heat of formation to an error of less than 1~meV/atom.
For the Co-based binaries we performed spinpolarised calculations with ferromagnetic 
and different antiferromagnetic configurations of the initial magnetic moments. 
In order to assess the relative stability of the different structures and stoichiometries we determined 
their respective heat of formation $\Delta H_f$ per atom,
\begin{equation} \label{eq:HoF}
 \Delta H_f =  \frac{E_{\mathrm{A-B}}-N_{\mathrm A} E_{\mathrm A} - N_{\mathrm B} E_{\mathrm B}}{N_{\mathrm{A-B}}}  \, ,
\end{equation}
where the indices A, B, A-B denote the number of atoms $N$ or the total energy $E$ of the elemental 
systems A, B and the compound system A-B, respectively. Our results for the TCP phases compiled in Ref.~\cite{Seiser-11-2}
are in good agreement with others, e.g. Ref.~\cite{Sluiter-07}. 
The calculated lattice constants are also in good agreement with experimental data and other DFT calculations 
as shown in Tab.~\ref{tab:lattice-constants}. 
\begin{table}
\caption{\label{tab:lattice-constants}
         Comparison of lattice parameters a$_0$ (and c/a ratio if applicable) of TCP
         phases as observed experimentally and in DFT calculations. The
         lattice parameters refer to those chemical compositions that are part of the convex hull 
         in our calculations (Figs.~\ref{fig:VNb-Ta} to~\ref{fig:VNb-Co}) and at the same time 
         close to or identical to the chemical compositions observed experimentally. 
         For Nb-Re we compare our results to other DFT calculations~\cite{Liu-13} for the most stable
         $\sigma$ phase composition at 2:1. For the Co-based compounds we provide in addition the 
         result of the nonmagnetic calculation in square brackets.}
\begin{indented}
\item[]\begin{tabular}{@{}lllll}
\br
\multicolumn{1}{l}{binary} & \multicolumn{1}{l}{TCP} & \multicolumn{1}{l}{A:B} & \multicolumn{2}{c}{a$_0$[\AA] /(c/a)}\\ 
\multicolumn{3}{l}{} & \multicolumn{1}{c}{DFT} & \multicolumn{1}{c}{expt~\cite{Landolt}} \\ 
\mr
V-Ta  & C14                   & 2:1  & 4.99 (1.59)                      & 4.96 (1.64)~\cite{Pearson}     \\ 
      &                       &      & 5.06 (1.59)~\cite{Pavlu-11}      &             \\  
      & C15                   & 2:1  & 6.99                             & 7.16        \\
      &                       &      & 7.10~\cite{Pavlu-11}             &             \\  
      & C36                   & 2:1  & 4.97 (3.21)                      & (not observed) \\
      &                       &      & 5.04 (3.20)~\cite{Pavlu-11}      &             \\  
\mr
V-Re  & A15                   & 1:3  & 4.85                             & 4.87        \\
      & $\sigma$              & 8:22 & 9.47 (0.50)                      & 9.44 (0.52) \\
\mr
Nb-Re & $\sigma$              & 2:3  & 9.68 (0.52)                      & 9.72 (0.52) \\
      &                       & 2:1  & 9.86 (0.52)                      &             \\
      &                       & 2:1  & 9.78 (0.52)~\cite{Liu-13}        &             \\
      & $\chi$                & 4:25 & 9.59                             & 9.78        \\
      &                       &      & 9.68~\cite{Liu-13}               &             \\  
\mr
Cr-Co & $\sigma$              & 2:1  & 8.49 (0.48)                      & 8.78 (0.52) \\
      &                       &      & [8.48 (0.51)]                    &             \\  
\mr
Mo-Co & $\sigma$              & 2:1  & 9.14 (0.52)                      & 9.23 (0.52) \\ 
      &                       &      & [9.14 (0.52)]                    &             \\ 
      & $\mu$                 & 6:7  & 4.66                             & 4.66  \\
      &                       &      & [4.66]                           &             \\
\mr
V-Co  & A15                   & 3:1  & 4.55                             & 4.68        \\ 
      &                       &      & [4.55]                           &             \\
      &                       &      & 4.67~\cite{Paduani-09}           &             \\  
      & $\sigma$              & 2:1  & 8.75 (0.51)                      & 9.03 (0.47) \\
      &                       &      & [8.76 (0.51)]                    &             \\ 
      & $\sigma$              & 1:2  & 8.41 (0.53)                      & 8.82 (0.46) \\
      &                       &      & [8.38 (0.54)]                    &             \\ 
\mr
Nb-Co & $\mu$                 & 7:6  & 4.79                             & 4.85  \\ 
      &                       &      & [4.78]                           &             \\ 
      & C14                   & 1:2  & 4.67 (1.63)                      & 4.83 (1.62) \\
      &                       &      & [4.67 (1.63)]                    &             \\ 
      & C15                   & 1:2  & 6.60                             & 6.77        \\
      &                       &      & [6.60]                           &             \\ 
      & C36                   & 1:2  & 4.67 (3.26)                      & 4.74 (3.21) \\
      &                       &      & [4.67 (3.26)]                    &             \\ 
\br
\end{tabular}
\end{indented}
\footnotetext[1] {metastable thin film}
\end{table}

\subsection{High-throughput environment}
\label{sec:HTE}

The DFT calculations were carried out within a high-throughput environment (HTE), an approach that is commonly 
used also in other groups, see e.g. Refs.~\cite{Curtarolo-05, Jain-11}. Our HTE \verb+strucscan+ automatically carries out
a large portion of the routine tasks, restarts interrupted or incomplete jobs, works in different computing environments, and organises
the storage of results for analysis.
\begin{figure}[htb]
\begin{center}
\includegraphics[width=0.6\textwidth]{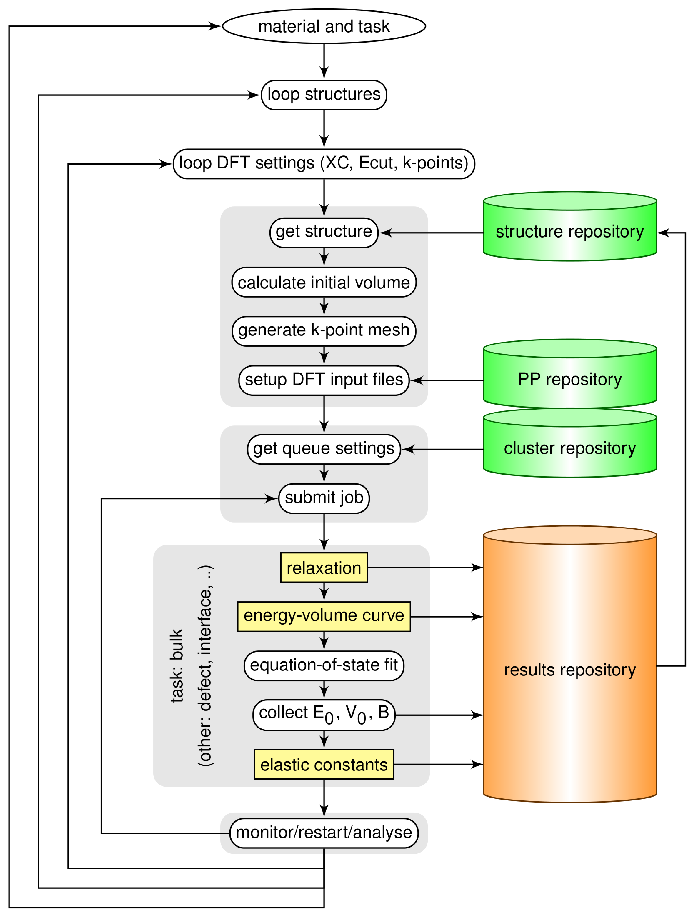}
\caption{The high-throughput environment for the DFT calculations queries repositories of crystal structures, 
pseudo-potentials and machine-configurations (green) and stores the output in a results repository (orange). 
The VASP calculations (yellow boxes) are carried out on available computing resources.}
\label{fig:strucscan}
\end{center}
\end{figure}
The input to the HTE defines a list of calculations that, for a given compound, loops over the specified exchange-correlation 
functionals, cutoff energies, $\mathrm k$-point mesh densities, initial magnetic moments and structures (Fig.~\ref{fig:strucscan}). 
For each combination of structure and DFT settings, the HTE carries out different tasks. 
Here we made use of determining the bulk ground-state only with a relaxation using an interpolated compound volume followed
by the calculation of the energy-volume curve which is fitted to the Murnaghan equation of state.
The HTE generates all input files, submits the job to the queueing system of the computing environment, monitors the status of the 
calculation and restarts it if required. 
Intermediate and final results are stored in a database to facilitate post-processing and restarts of interrupted calculations. 
Our HTE has proven to be stable in earlier work on intermetallics~\cite{Hammerschmidt-08-2,Seiser-11-2}, elemental transition 
metals~\cite{Qin-08}, steels~\cite{Kolmogorov-10,Bialon-11,Bialon-13} and battery materials~\cite{Hajiyani-13}.

\section{Results and discussion}
\label{sec:results}

\subsection{Role of size at $\Delta N =0$: V/Nb-Ta}
\label{sec:VNb-Ta}

We first discuss V-Ta and Nb-Ta, two isovalent systems ($\Delta N =0$) at $N=5$ and constituent elements with bcc ground states. 
The observed experimental phases~\cite{Landolt} are bcc solid solutions with an additional C15 phase for $T<1600 \mathrm{K}$
at a chemical composition of approximately 2:1 in V-Ta. 
According to the range of the average valence-electron count $\Delta N=0$, the Nb-Ta and the V-Ta 
system are represented by line-segments at ${\bar N}=5$ in the structure map of Fig.~\ref{fig:strucmap}. 
The small value of the relative volume difference of $\overline{\Delta V/V}$=0.001 for Nb-Ta (cf. Tab.~\ref{tab:systems}) 
confines the line segment to nearly a point. The value of $\overline{\Delta V/V}$=0.128 for V-Ta results in a line segment 
that intersects with the region of Laves phase stability as shown in Fig.~\ref{fig:strucmap}. 
Hence, we expect no TCP phase at all in Nb-Ta but possibly one of the Laves phases in V-Ta, in agreement with experiment. 

\begin{figure}[htb]
\begin{center}
\includegraphics[width=0.49\textwidth,angle=0]{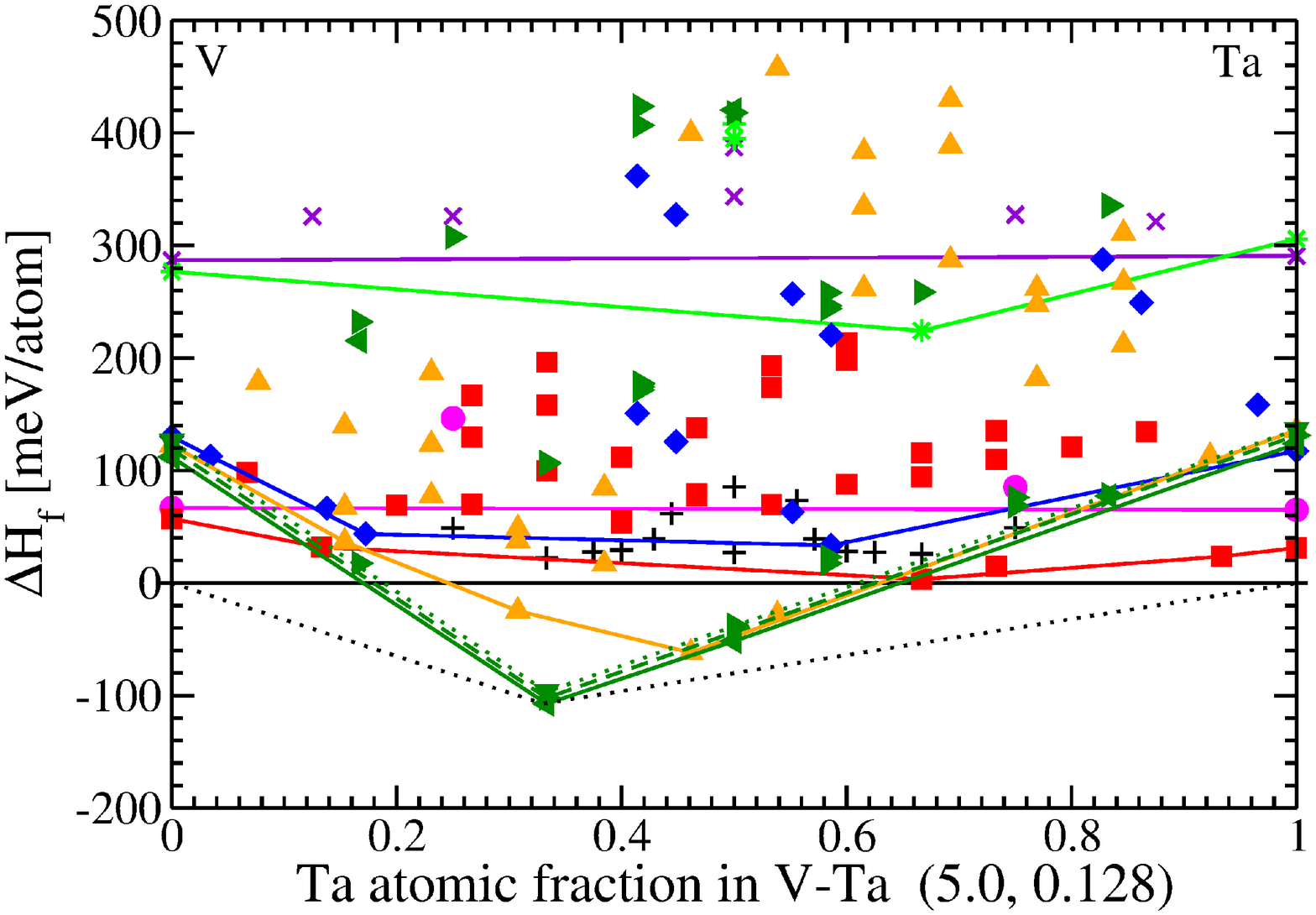}
\includegraphics[width=0.49\textwidth,angle=0]{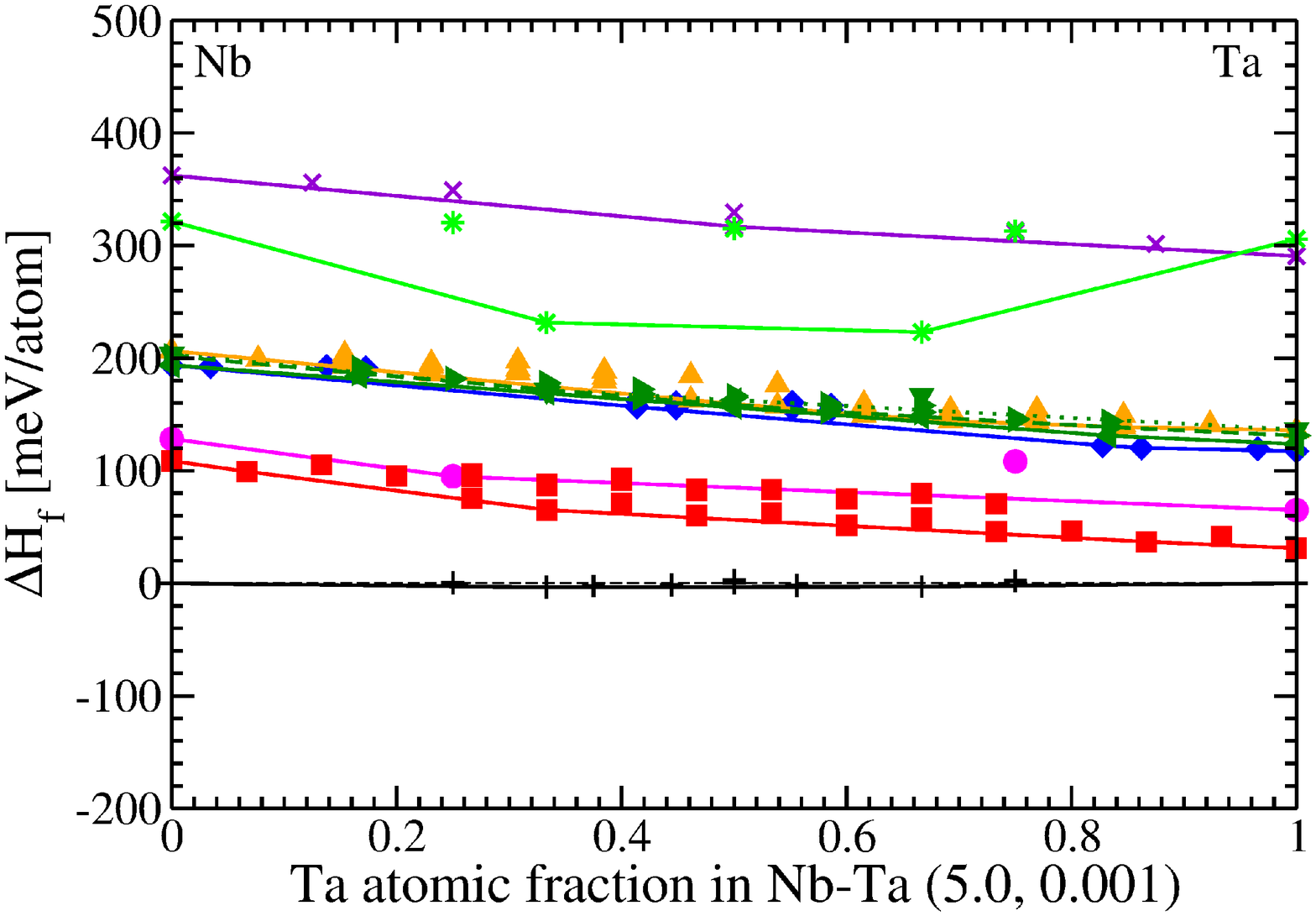}\\
\includegraphics[width=0.45\textwidth,angle=0]{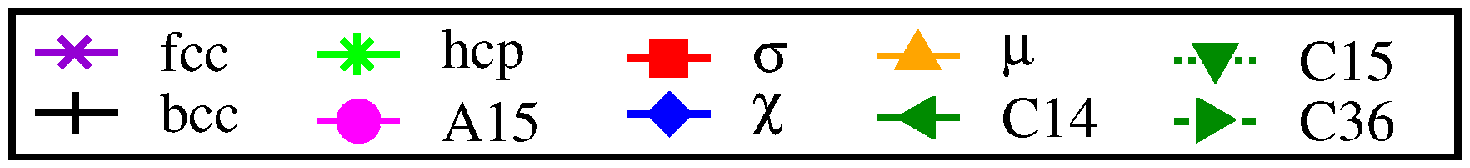}
\caption{Heat of formation of hcp, fcc, bcc and TCP phases from DFT calculations for V-Ta
         and Nb-Ta. In the structure map, this pair of systems corresponds to a comparison at 
         $\Delta N =0$ at different $\overline{\Delta V/V}$ and hence illustrates the role of 
         atomic size that stabilises the $\mu$ phase (orange) and the Laves phases (green) in this 
         isovalent comparison.
         The structure map coordinates ($\overline{N}_{x=0.5}$, $\overline{\Delta V/V}$) are given in brackets.}
\label{fig:VNb-Ta}
\end{center}
\end{figure}

For the comparison with DFT calculations, we compiled the heat of formation of the considered structures (Tab.~\ref{tab:prototypes}) 
with respect to the elemental ground states (Eq.~\ref{eq:HoF}) in Fig.~\ref{fig:VNb-Ta}.
The set of data points that we obtain by permutating the two constituent elements on the Wyckoff positions are 
shown in the same colour, together with the convex hull of the heat of formation for each crystal structure.
Each Wyckoff position contributes a set of atoms to the unit cell according to its multiplicity. 
If the same chemical composition can be achieved with different combinations of Wyckoff occupancies we also find
multiple entries of the heat of formation for the same chemical composition. For example for the $\mu$ phase, the
Wyckoff positions have the multiplicities 1/6/2/2/2 and hence lead to two realisations A/B/A/A/A and A/A/B/B/B
of a 7:6 stoichiometry in an A-B compound. 
For Nb-Ta we find an approximately linear relation between the heat of formation and the chemical composition. This is not surprising 
given the identical valence-electron count and similar atomic size. We 
find bcc-based solid-solutions to be most stable throughout the whole range of composition, in line with the 
experimental phase diagram. 
For V-Ta we also find a linear relation for the heat of formation of bcc, fcc, hcp, A15, $\chi$ and $\sigma$ phases.
The $\mu$ phase and the Laves phases C14, C15, C36 show a different behaviour with a pronounced minimum of
the convex hull at V$_2$Ta (Laves) and at V$_7$Ta$_6$ ($\mu$). 
The observation of a Laves phase at a composition of V$_2$Ta is consistent with experiment. In our 
calculations, we observe C14 as the most stable phase, followed by C36 and C15. This difference to the 
experimentally observed C15 phase can be attributed to the closely competing energies of the Laves phases 
of only a few meV/atom due to their structural similarity. From our DFT results we would also expect the $\mu$ 
phase and the $\sigma$ phase to be nearly stable at approximately 1:1 and 1:2 composition, respectively.

Juxtaposing our DFT results for the two compounds, we observe that the bcc, fcc, hcp, A15, $\chi$ and 
$\sigma$ phases are very similar in this comparison at constant valence-electron count, even 
for different size factors. The $\mu$ and Laves phases, however, are sensitive to differences in the 
atomic size: these phases are considerably more stable in our DFT calculations for V-Ta (upper panel 
of Fig.~\ref{fig:VNb-Ta}) with a larger value of $\overline{\Delta V/V}$, at constant valence-electron 
count. This observation is in line with earlier work that identified the grouping of TCP phases
in these distinct sets~\cite{Hammerschmidt-08-2,Seiser-11-2}.

\subsection{Role of size at $\Delta N$=2: V/Nb-Re}
\label{sec:VNb-Re}

In V-Re and Nb-Re the values of $\overline{\Delta V/V}$ are 0.0320 and
0.0979, respectively (cf. Tab.~\ref{tab:systems}), at a variation in average valence-electron 
count of $\Delta N=2$. The elemental ground states are bcc (V, Nb) and hcp (Re). 
The experimental phase-diagrams show narrow stability ranges of the $\sigma$
phase at approximately V$_{24}$Re$_{76}$ and Nb$_{45}$Re$_{55}$. Furthermore, both systems show
a stable $\chi$ phase around V$_{29}$Re$_{71}$ and for approximately 63-87~at.\%~Re in Nb-Re. 
In the structure map (Fig.~\ref{fig:strucmap}), the parabola of these systems range from ${\bar N}=5$ 
to ${\bar N}=7$ and cross the regions of A15, $\sigma$ and $\chi$.  

\begin{figure}[htb]
\begin{center}
\includegraphics[width=0.49\textwidth,angle=0]{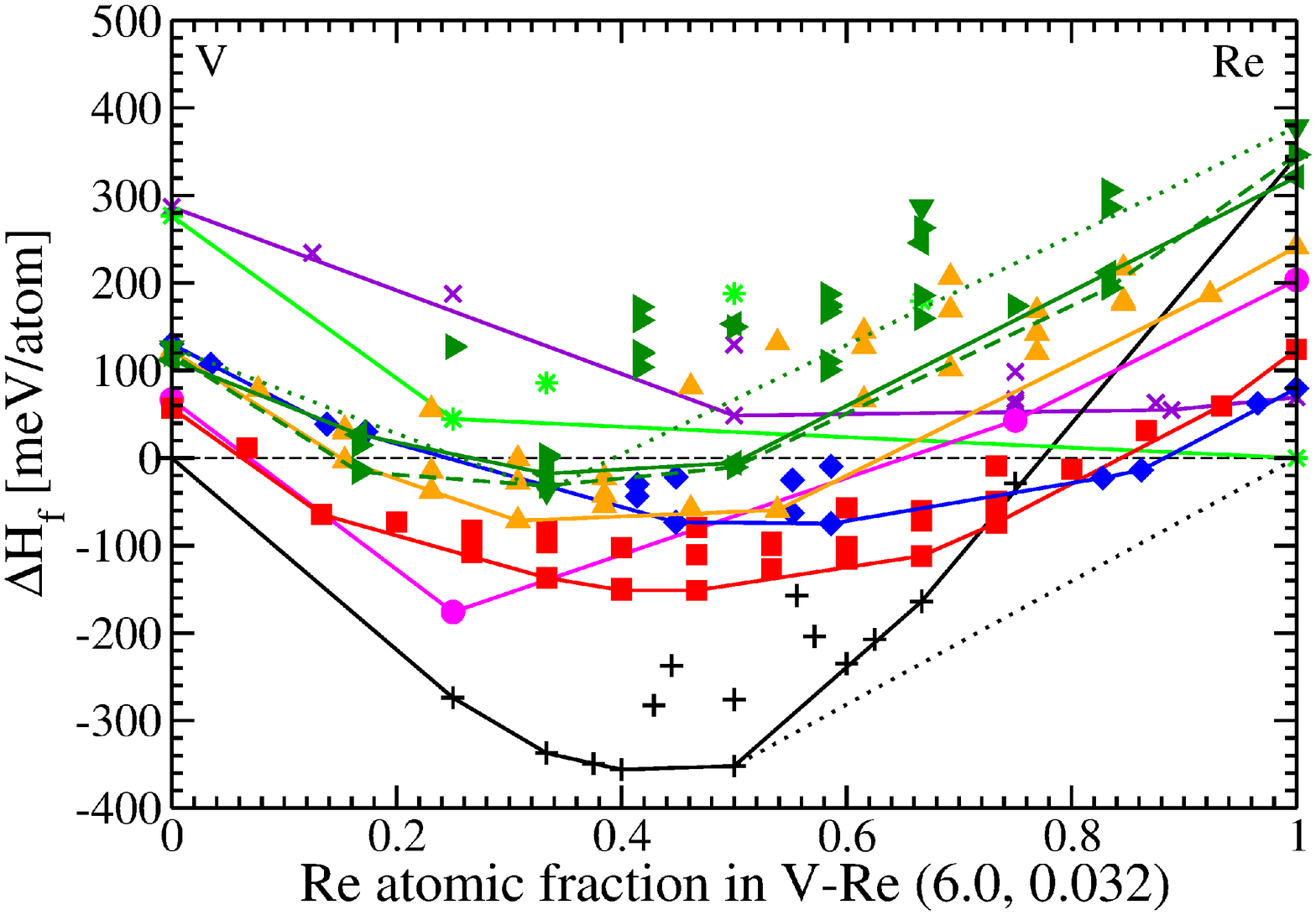}
\includegraphics[width=0.49\textwidth,angle=0]{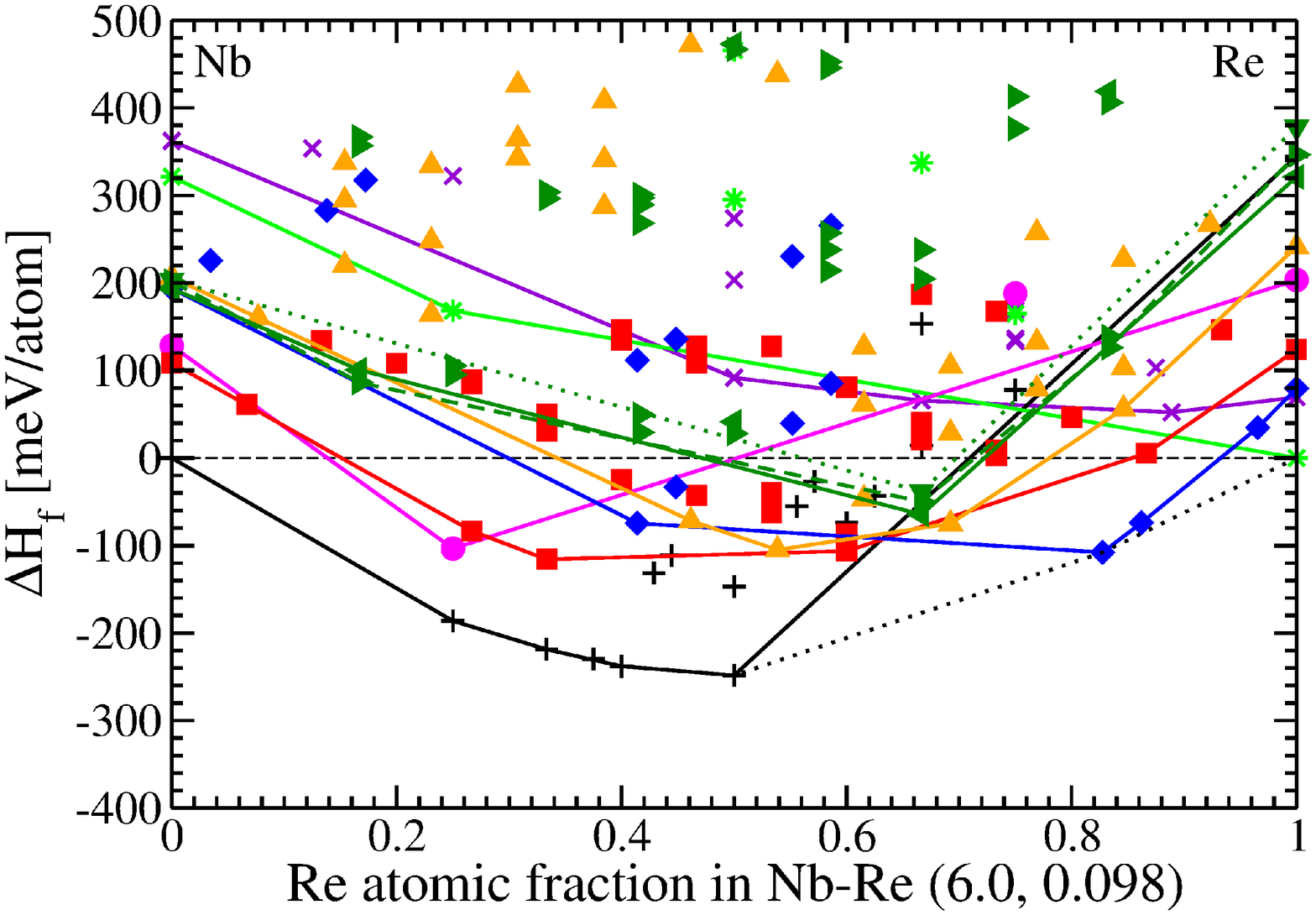}\\
\includegraphics[width=0.45\textwidth,angle=0]{fig3_6-legend.eps}
\caption{Heat of formation of hcp, fcc, bcc and TCP phases from DFT calculations 
         for V-Re and Nb-Re. In the structure map, this pair of systems corresponds to a
         comparison at $\Delta N =2$ at different $\overline{\Delta V/V}$ and shows
         the influence of $\bar{N}$ on the structural sequence A15, $\sigma$, and $\chi$. 
         The structure map coordinates ($\overline{N}_{x=0.5}$, $\overline{\Delta V/V}$) are given in brackets.}
\label{fig:VNb-Re}
\end{center}
\end{figure}

This is consistent with our DFT results compiled in Fig.~\ref{fig:VNb-Re}
where the minima of the TCP phases follow the trend A15$\rightarrow \sigma \rightarrow \chi$
with increasing $\bar{N}$ from V or Nb to Re. 
Part of the results for V-Re was used in an assessment of the influence of the DFT settings 
on the construction of finite-temperature phase diagrams from DFT results~\cite{Palumbo-13}.
All TCP phases 
show a range of chemical composition with negative heat of formation in both systems, in contrast
to, e.g., Mo-Re~\cite{Seiser-11-2} with $\overline{\Delta V/V}$ similar to Nb-Re at $\Delta N=1$.
However, the stabilisation of TCP phases is superseded by the bcc phase that 
is even more stable for a range of about 0-70~at.\% Re in V-Re and for 0-60~at.\% Re in Nb-Re. 
In V-Re, we observe narrow composition ranges of a stable $\sigma$ phase near 0.75~at.\%~Re and
a stable $\chi$ phase near 0.85~at.\%~Re. While the experimentally observed $\sigma$ phase has a
similar composition, the $\chi$ phase was found at a lower concentration of Re of 0.70~at.\%.
This discrepancy might be due to entropy, as according to our DFT data the 
$\sigma$ and the $\chi$ phase are closely competing with the bcc phase in this range of composition
while these phases are observed experimentally only for $T>1700 \mathrm{K}$. The broad range
of stability of the $\chi$ phase in Nb-Re from approximately 65~at.\% to 90~at.\%~Re in our results
is in very good agreement with the experimental phase diagram. 
In comparison to V/Nb-Ta we find that the smaller difference of $\overline{\Delta V/V}$ in V/Nb-Re
destabilises the $\mu$ phase and the Laves phases in V-Re.
Also we find that the position of the $\mu$ and Laves phase minima changes from V$_2$Re to NbRe$_2$. 
This change of occupancy of the crystallographic sublattices of the Re atoms from the CN16 polyhedra 
to the smaller CN12 polyhedra is a consequence of being the larger atoms in V-Re but the smaller ones
in Nb-Re.
Our results for V-Re and Nb-Re are also consistent with a recent DFT screening of the $\chi$ and $\sigma$
phase in Re-based binary TM compounds~\cite{Crivello-13}.

\subsection{Role of magnetism and size at $\Delta N$=3: Cr/Mo-Co}
\label{sec:CrMo-Co}

The binaries Cr-Co and Mo-Co illustrate the role of $\bar{N}$ and $\overline{\Delta V/V}$ in the
presence of magnetism. For Mo and Cr the ground state structure is bcc, for Co the ground state 
undergoes a transition from hcp ($\varepsilon$-Co) to fcc ($\alpha$-Co) near T=695~K at ambient pressure
(see e.g. Ref.~\cite{Toledano-01} and references therein). 
In Cr-Co, the phase diagram shows the $\sigma$ phase in a broad composition range of 32-47~at.\%~Co for 
$800 \mathrm{K}<T<1556 \mathrm{K}$. In Mo-Co, the $\sigma$ phase is stable only in a narrow composition 
range near Mo$_3$Co$_2$ and for $1270 \mathrm{K}<T<1893 \mathrm{K}$. 
In addition, the Mo-Co system shows a region of a stable $\mu$ phase at around 52-58~at.\%~Co, as well as 
ordered hexagonal structures at MoCo$_3$ (D0$_{19}$) and MoCo$_4$. 
The stability of the TCP phases is consistent with the results of the structure map where Cr-Co and Mo-Co are represented 
by parabola between $N=6$ and $N=8$ (Fig.~\ref{fig:strucmap}). Both systems show intersections with the regions 
of A15, $\chi$ and $\sigma$. For Co-rich compositions the structure map suggests that no TCP phase 
forms, in agreement with the experimental phase diagrams. In both systems, the composition of the experimentally observed 
$\sigma$ phase is in line with the structure map, while no A15 or $\chi$ phase has been reported in these binaries. 
For Cr-Co, the $\mu$ and Laves phases are neither apparent in the experimental phase diagram nor 
expected from the structure map due to the small value of $\overline{\Delta V/V}$ of 0.0355. For Mo-Co, 
however, the larger value of $\overline{\Delta V/V}$ of 0.1660 leads to additional intersections with 
the region of $\mu$ and Laves phase stability that match the experimentally observed $\mu$ phase
just below 1:1 composition.

\begin{figure}[htb]
\begin{center}
\includegraphics[width=0.49\textwidth,angle=0]{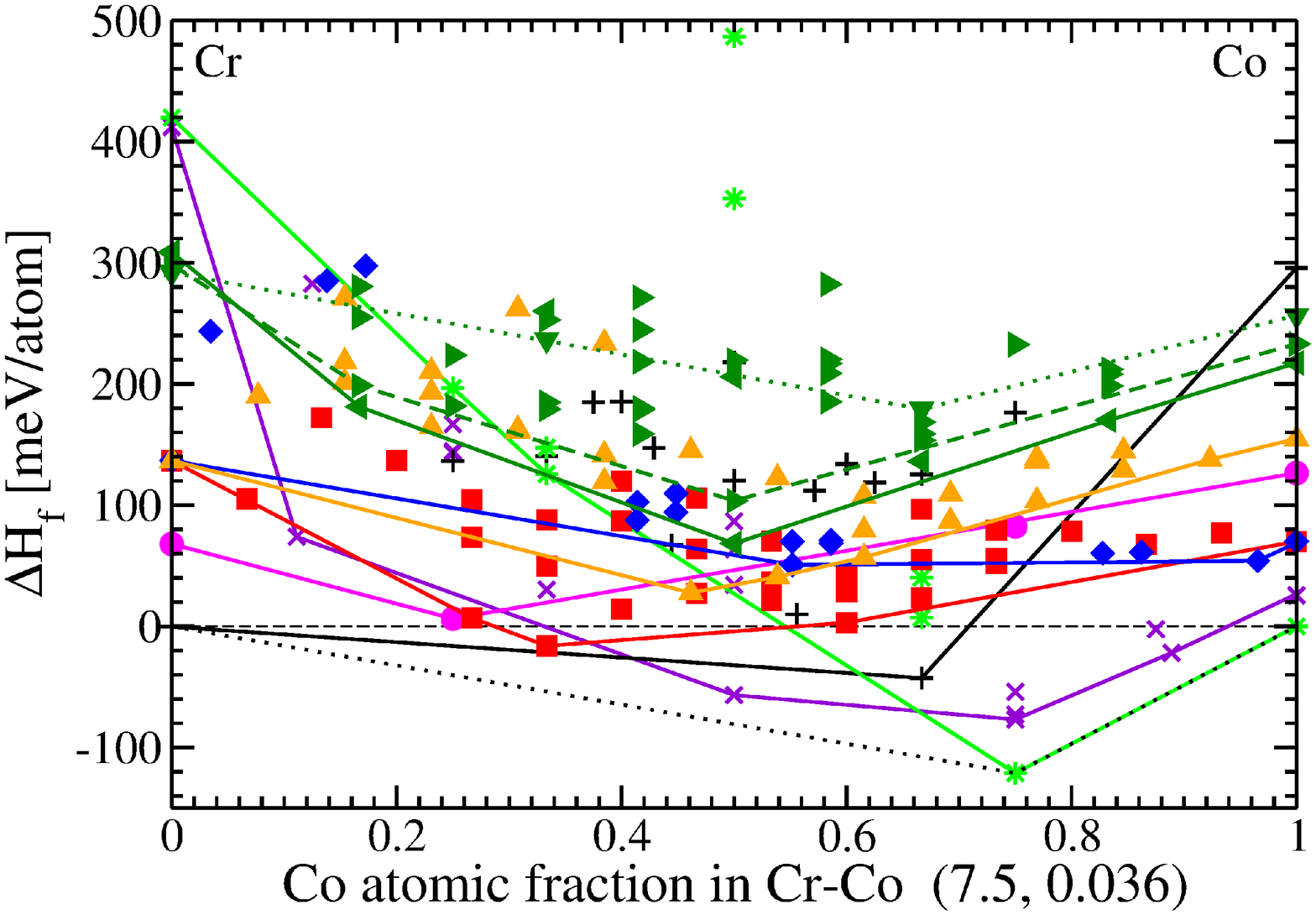}
\includegraphics[width=0.49\textwidth,angle=0]{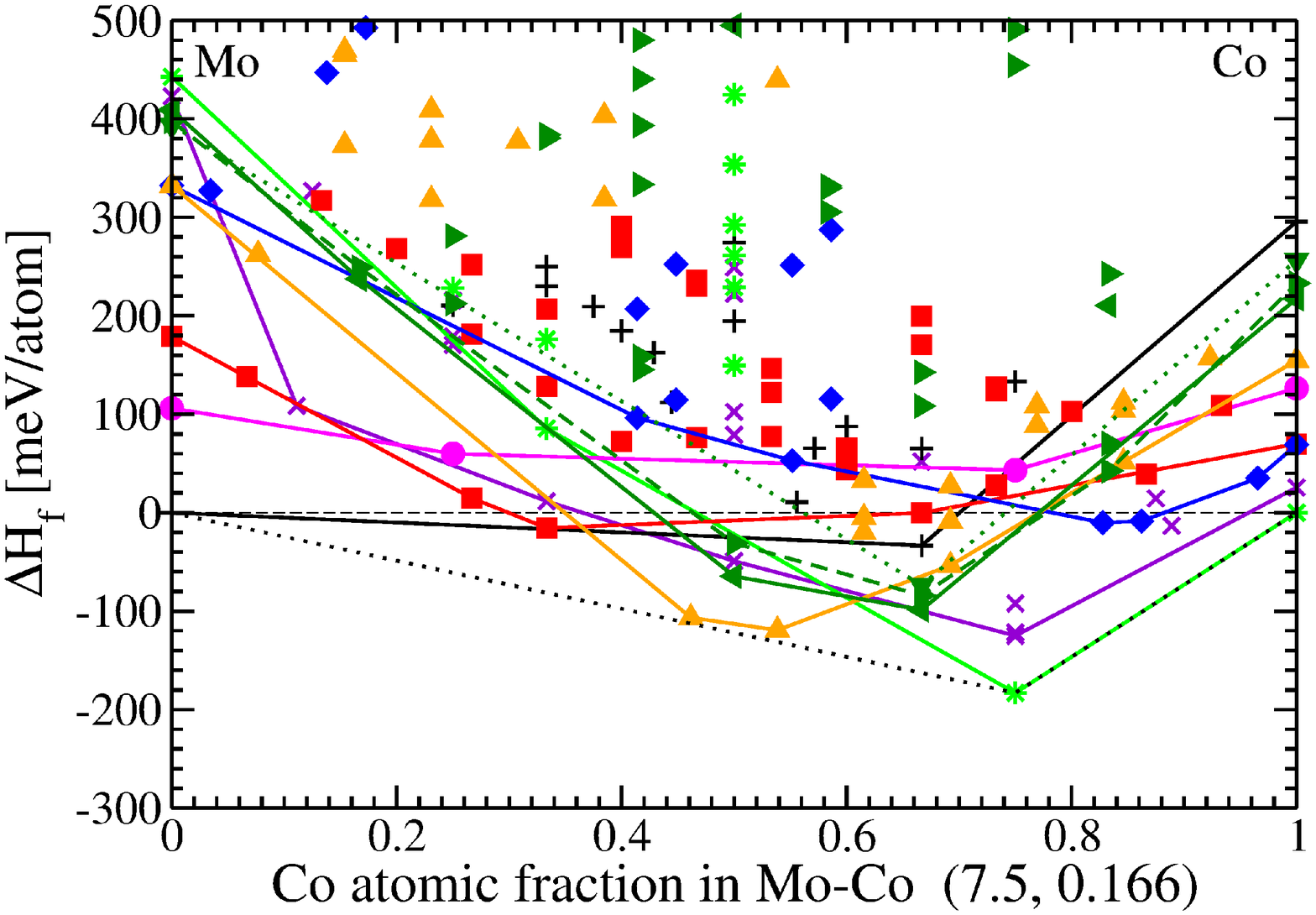}\\
\includegraphics[width=0.45\textwidth,angle=0]{fig3_6-legend.eps}
\caption{Heat of formation of hcp, fcc, bcc and TCP phases from DFT calculations 
         for Cr-Co and Mo-Co. In the structure map, this pair of systems corresponds to a
         comparison at $\Delta N =3$ at different $\overline{\Delta V/V}$ and shows
         the influence of atomic size differences. 
         The structure map coordinates ($\overline{N}_{x=0.5}$, $\overline{\Delta V/V}$) are given in brackets.}
\label{fig:CrMo-Co}
\end{center}
\end{figure}

For the Cr-Co and Mo-Co systems we carried out two sets of DFT calculations: spinpolarised (Fig.~\ref{fig:CrMo-Co}) and non-spinpolarised.
We find that the major difference between the two sets relates to the ground-state of Co that is fcc in the non-spinpolarised
and, in agreement with experiment, hcp in the spinpolarised calculations. 
In general we also observe an increasing average magnetic moment with increasing Co content in line with previous findings,
e.g., for the $\sigma$ phase in Cr-Co~\cite{Pavlu-10}.
By comparing our spinpolarised and non-spinpolarised DFT calculations we find that the influence of magnetism on the 
heat of formation and on the average magnetic moments of TCP phases in the convex hull are sizeable only for more than 
approximately 75~at.\%~Co. However, in this Co-rich range the TCP phases are neither stable in our DFT calculations nor 
expected from the structure map. 

Comparing our DFT results for Cr-Co with the phase diagram we find that the experimentally observed 
broad composition range for the $\sigma$ phase near Cr$_2$Co is in line with the slightly negative
heat of formation that we observe for this composition. 
According to the DFT results and the structure map one might expect to observe the A15 phase near Cr$_3$Co. 
The $\mu$ and Laves phases are higher in energy as we would also expect due to a small value of  $\overline{\Delta V/V}$ of 0.0355.
For the Mo-Co system, the phases A15, $\sigma$ and $\chi$ exhibit similar heat of formations as
expected for this isovalent comparison of binaries, only the A15 phase at 3:1 composition is slightly
less stable in Mo-Co. The large value of $\overline{\Delta V/V}$ of 0.1660, stabilises the $\mu$ and Laves 
phases by about 200~meV/atom at around 1:2 and 1:2 compositions, respectively. 
This is consistent with the experimental phase diagram that reports a $\sigma$ phase at Mo$_3$Co$_2$ and a broad range of $\mu$ stability 
around 1:1 compositions. 
Both systems exhibit a D0$_{19}$ structure that we find to be most stable at CrCo$_3$ and MoCo$_3$.

\subsection{Role of magnetism and size at $\Delta N$=4: V/Nb-Co}
\label{sec:VNb-Co}

In order to investigate the role of $\overline{\Delta V/V}$ in combination with a large 
variation of $\bar{N}$, we compare the binaries V-Co and Nb-Co. 
Among the binaries investigated here, V-Co is the only system with a stable A15 phase (at V$_3$Co), 
although all considered Re-based and Co-based binaries intersect the A15 region in the structure map. 
Additionally, the V-Co system forms a $\sigma$ phase for 30-55~at.\%~Co and exhibits a D0$_{19}$ $\rightarrow$ L1$_2$ phase transition at VCo$_3$. 
The Nb-Co system shows the full set of size-determined TCP phases, ranging from $\mu$-Nb$_6$Co$_7$ over C14-NbCo$_2$ and 
C15-NbCo$_2$ to C36-NbCo$_3$. The phase transition of C14-NbCo$_2$ to C15-NbCo$_2$ at $T>1474 \mathrm{K}$ 
proposed earlier~\cite{Landolt} was refined experimentally~\cite{Stein-08} to separate off-stoichiometric 
stability regions of C14 (63-64~at.\% Co) and C15 (65-74~at.\% Co). The coexistence of different 
off-stoichiometric Laves phases can be attributed to their different flexibility for accommodating anti-site 
defects~\cite{Stein-11}. 
The experimental refinement~\cite{Stein-08} also ruled out the existence of previously reported L1$_2$ precipitates~\cite{Landolt} 
and reported an additional Nb$_2$Co$_7$ phase. This phase was not included in the investigation of TCP phases presented here and
is expected to be more stable than D0$_{19}$ and L1$_2$.
The full set of TCP phases in these two systems is also apparent in the structure map that shows intersections with the regions of
A15, $\sigma$ and $\chi$ in the case of V-Co and with A15, Laves and $\mu$ in the case of Nb-Co. 
As for Cr/Mo-Co, we expect no formation of TCP phase in V/Nb-Co for more than approximately 75~at.\%~Co, in line with experiment. 

\begin{figure}[htb]
\begin{center}
\includegraphics[width=0.49\textwidth,angle=0]{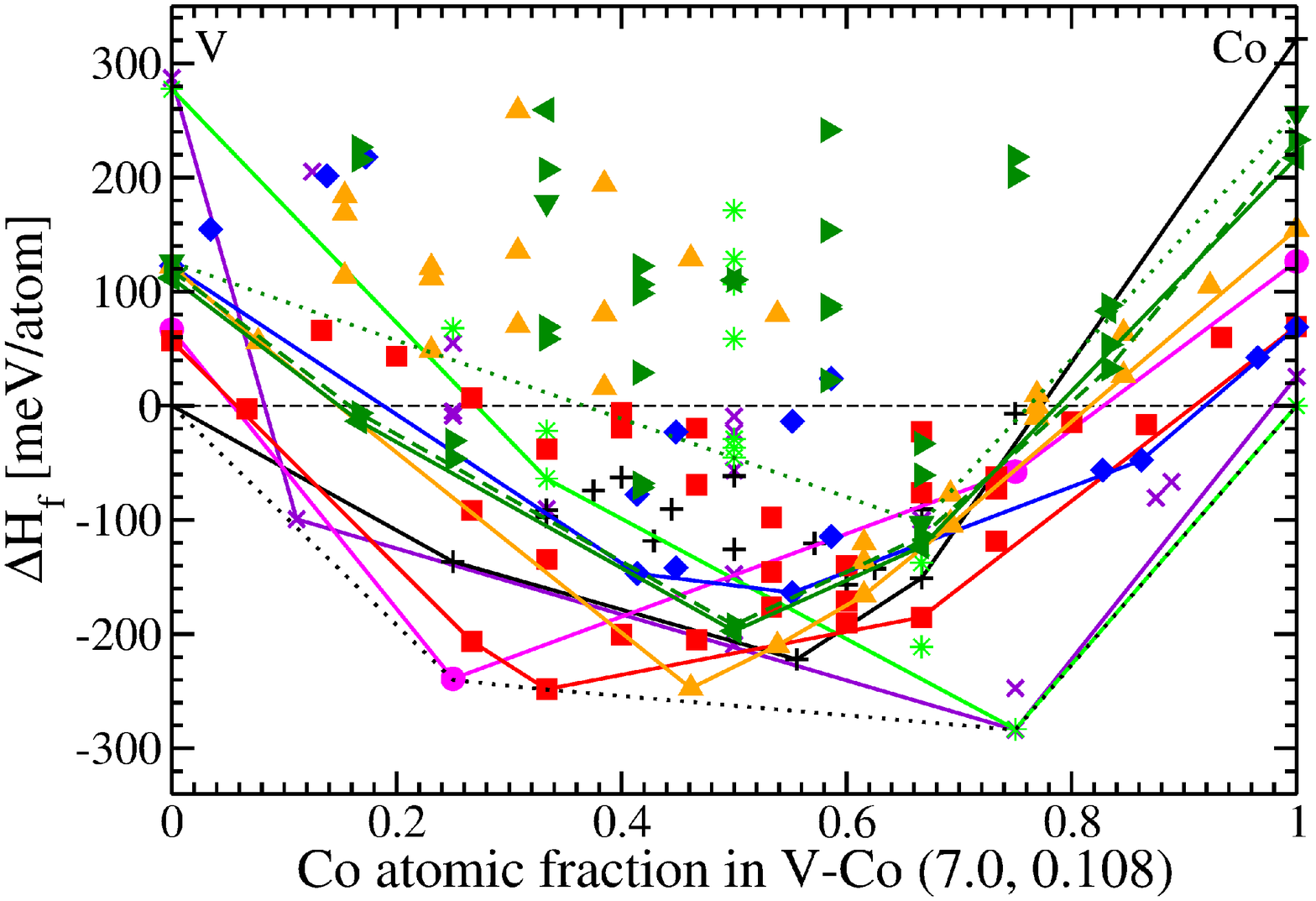}
\includegraphics[width=0.49\textwidth,angle=0]{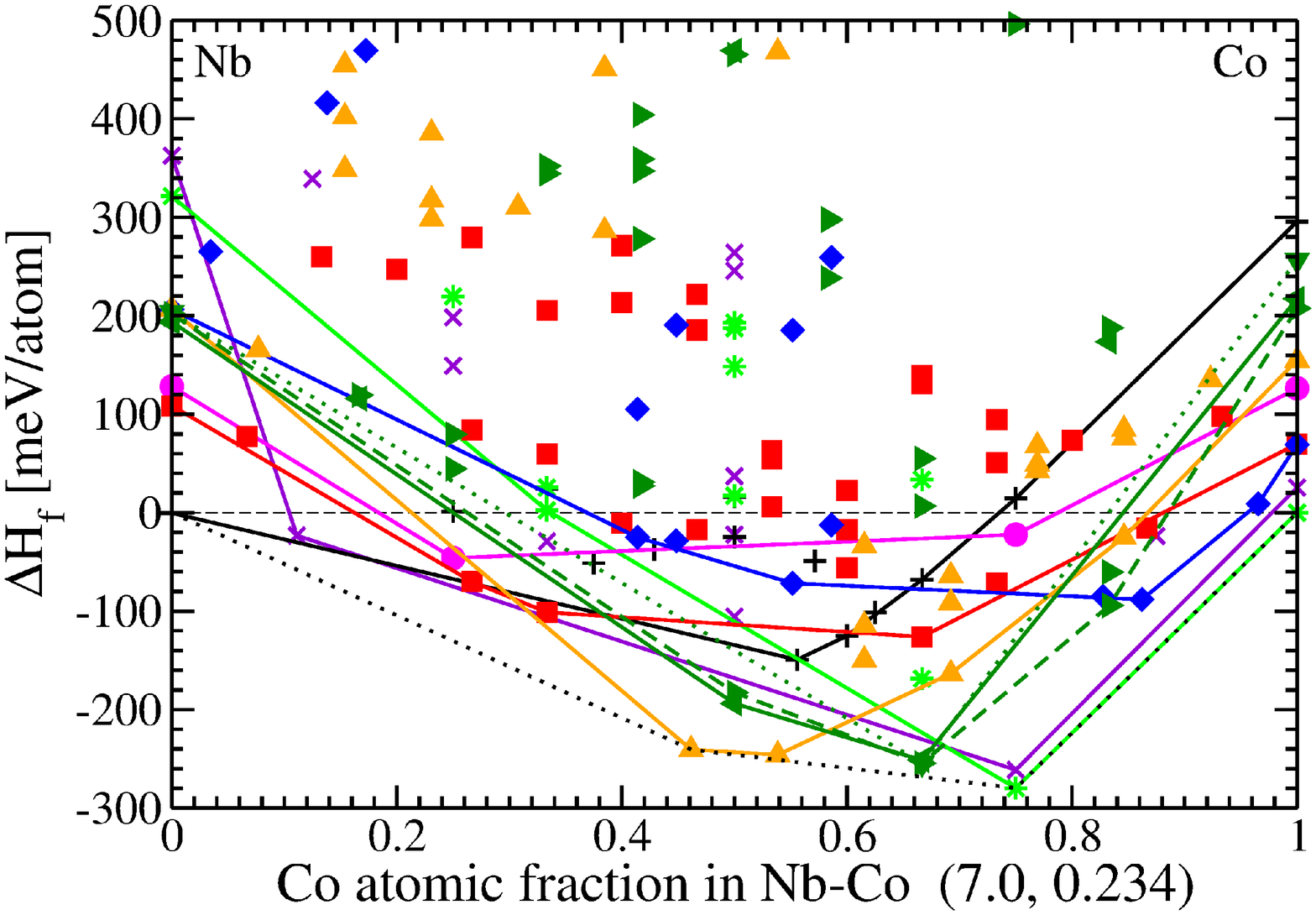}\\
\includegraphics[width=0.45\textwidth,angle=0]{fig3_6-legend.eps}
\caption{Heat of formation of hcp, fcc, bcc and TCP phases from DFT calculations 
         for V-Co and Nb-Co. In the structure map, this pair of systems corresponds to a
         comparison at $\Delta N =4$ at different $\overline{\Delta V/V}$ and shows
         the influence of atomic size differences. 
         The structure map coordinates ($\overline{N}_{x=0.5}$, $\overline{\Delta V/V}$) are given in brackets.}
\label{fig:VNb-Co}
\end{center}
\end{figure}

The DFT calculations were carried out in two sets, spinpolarised (shown in Fig.~\ref{fig:VNb-Co}) and 
non-spinpolarised as described for Cr/Mo-Co above. Similar to those systems, we find an influence of magnetism on the phase stability in V-Co and Nb-Co only for Co 
concentrations above approximately 75~at.\%. This may again be attributed to the small values of the average magnetic 
moments for those structures close to the convex hull with less than about 75~at.\% Co.
In our DFT results we find a close competition of the stable L1$_2$ and the hexagonal 
D0$_{19}$ phase at VCo$_2$ and at NbCo$_2$, in line with the experimental phase diagram and in 
good agreement with previous ab-initio calculations~\cite{Lin-92}. 

We also find compositions with negative values of the heat of formation for all TCP phases in
both systems. The compositions of the stable A15-V$_3$Co, $\sigma$-V$_2$Co, $\mu$-Nb$_6$Co$_7$ to 
$\mu$-Nb$_7$Co$_6$, as well as C14-NbCo$_2$ and C15-NbCo$_2$ match the experimental phase diagrams.
We find the reported C36-NbCo$_3$ phase too high in energy to play a role in the phase diagram (assuming a ferromagnetic state for both 
sublattice arrangements that lead to a 1:3 composition). However, we find a C36-Nb$_4$Co$_{20}$ phase that closely competes with $\chi$-Nb$_5$Co$_{20}$
and $\chi$-Nb$_4$Co$_{25}$, but is located about 100~meV/atom above the convex hull of the most stable
hcp structures. We attribute this discrepancy to the neglect of substitutional defects in our study 
that have been identified to stabilise the C36-NbCo$_3$ in experimental phase-diagrams~\cite{Stein-08}.

\section{Conclusions}
\label{sec:conclusion}

By pairwise comparisons of binary systems we show that DFT results for TCP phases may be understood in terms of an empirical structure-map.
This demonstrates that the average valence-electron count $\bar{N}$ and the relative volume-differences $\overline{\Delta V/V}$ can
be employed to rationalise the structural stability of TCP phases in order to support the selection of compounds for materials design.

In particular, we investigated the structural stability of the topologically close-packed phases A15, $\sigma$, 
$\chi$, $\mu$, C14, C15, and C36 in the binary transition-metal compounds V/Nb-Ta, V/Nb-Re and Cr/Mo/V/Nb-Co.
Using a high-throughput environment for DFT calculations, we determined the formation energy of the TCP 
phases in these binary systems for all occupations of the Wyckoff sites across the full range of 
chemical compositions. Overall, we find consistently the group of TCP phases of A15$\rightarrow \sigma \rightarrow \chi$ with 
increasing $\bar{N}$ for all systems, clearly separated from the group of $\mu$, C14, C15 and C36 which are stabilised by 
large values of $\overline{\Delta V/V}$. This is in line with the structure map and with previous insights from coarse-grained 
electronic structure methods. For the different systems we find in particular: 
\begin{itemize}

\item The comparison of the isovalent binaries (i.e. $\Delta N=0$) V-Ta and Nb-Ta show that the larger value of 
$\overline{\Delta V/V}$ for V-Ta stabilises the $\mu$ and Laves phases significantly to a negative value of the 
heat of formation, in contrast to Nb-Ta. In experiment, only the C15 phase has been observed.

\item The systems V-Re and Nb-Re with a larger $\Delta N=2$ shows a similar trend in the minima of
A15$\rightarrow \sigma \rightarrow \chi$ with $\bar{N}$. The stabilisation of the 
$\mu$ and Laves phases and the shift of their convex-hull minimum in Nb-Re can also be attributed 
to the larger $\overline{\Delta V/V}$. We find that all TCP phases in 
V-Re and in Nb-Re exhibit a composition range with negative formation energy. The stability of the 
$\chi$ phase observed in our DFT data is consistent with experimental phase-diagrams although the 
Re concentration there is slightly smaller.

\item In the systems Cr-Co and Mo-Co the large range of $\bar{N}$ from 6 to 9 leads to TCP formation only
for less than about 60~at.\% Co. This finding is consistently observed in the DFT calculations, the 
structure map and the experimental phase diagrams. The A15, $\sigma$ and $\chi$ phases show the same
trend with average valence-electron count, but only the $\sigma$ phase is sufficiently close to
the convex hull at concentrations that are in good agreement with the experimentally observed $\sigma$
phase compositions. The larger $\overline{\Delta V/V}$ in Mo-Co leads to a significant stabilisation of 
the $\mu$ and Laves phases in particular at 1:1 and 1:2 composition. The chemical range of $\mu$ phase
stability is in good agreement with the structure map and the experimental phase diagram. 

\item The V-Co and Nb-Co systems with a range of $\bar{N}$ from 5 to 9 show the largest amount
of TCP phases with negative formation energies. While the group of A15 and $\sigma$ phases are found
consistently in Co-V, the larger $\overline{\Delta V/V}$ in Nb-Co gives rise to the stabilisation of the
group of $\mu$/Laves phases. The 1:2 composition of stable C14 and C15 phases observed in our DFT
calculations is in agreement with experiment. The C36 phase at NbCo$_3$ known from phase diagrams is 
not found as we did not consider substitutional defects that stabilise C36 at this composition.

\end{itemize}

According to the average magnetic moments and to comparisons with additional non-spinpolarised 
calculations, we find only a weak influence of magnetism on the structural stability of TCP phases 
in Cr/Mo-Co and in V/Nb-Co.

\ack

We acknowledge discussions with M.~Palumbo, S.G.~Fries and J.-C.~Crivello. 
Part of this work was funded by the \emph{Alloys By Design} consortium of the Engineering and 
Physical Sciences Research Council (EPSRC) of the United Kingdom. 
We are grateful to our collaboration partners in this consortium, in particular R.C.~Reed, 
C.M.~Rae and N.~Warnken. 
We acknowledge financial support through ThyssenKrupp AG, Bayer MaterialScience AG, 
Salzgitter Mannesmann Forschung GmbH, Robert Bosch GmbH, Benteler Stahl/Rohr GmbH, Bayer
Technology Services GmbH and the state of North-Rhine Westphalia as well as the EU in the 
framework of the ERDF.
Part of this work was supported by the DFG through project C1 of SFB/TR 103.

\section*{Bibliography}

\bibliography{NJP-15-115016-2013}

\end{document}